%% file: sgracolor.tex
\documentclass{emulateapj}
\usepackage{apjfonts}
\usepackage{lscape}
\usepackage{appendix}
\slugcomment{To appear in the Astrophysical Journal, 1 Oct 2007, v667n2} 

\newcommand{\msun}{M$_{\sun}$\,}
\newcommand{\Ms}{$M_{s}$}
\newcommand{\Lp}{$L^\prime$}
\newcommand{\Kp}{$K^\prime$}
\newcommand{\Hp}{$H$}

\shorttitle{Constant Spectral Index for Sgr A*}
\shortauthors{Hornstein et al. }

\begin{document}

\title{A Constant Spectral Index for Sagittarius A* During
Infrared/X-ray Intensity Variations}

\author{S.~D. Hornstein\altaffilmark{1,2},
K. Matthews\altaffilmark{3},
A.~M. Ghez\altaffilmark{1,4}, 
J.~R. Lu\altaffilmark{1}, 
M. Morris\altaffilmark{1}, 
E.~E. Becklin\altaffilmark{1},
M. Rafelski\altaffilmark{1},
and F.~K. Baganoff\altaffilmark{5}
}
\email{seth.hornstein@colorado.edu}
\altaffiltext{1}{Department of Physics and Astronomy,
University of California, Los Angeles, CA 90095-1547}
\altaffiltext{2}{Current Address: Center for Astrophysics and Space
  Astronomy, Department of Astrophysical and Planetary Sciences, 
  University of Colorado, Boulder, CO 80309}
\altaffiltext{3}{Caltech Optical Observatories, California
Institute of Technology, MS 320-47, Pasadena, CA 91125}
\altaffiltext{4}{Institute for Geophysics and Planetary
Physics, University of California, Los Angeles, CA 90095-1565}
\altaffiltext{5}{Kavli Institute for Astrophysics and Space
Research, Massachusetts Institute of Technology, Cambridge, MA
02139-4307}

\begin{abstract}
We report the first time-series of broadband infrared color
measurements of Sgr A*, the variable emission source
associated with the supermassive black hole at the Galactic
Center.  Using the laser and natural guide star adaptive
optics systems on the Keck II telescope, we imaged Sgr A* in
multiple near-infrared broadband filters with a typical cycle
time of $\sim$3 min during 4 observing runs (2005--2006), two
of which were simultaneous with \textit{Chandra} X-ray
measurements.  In spite of the large range of dereddened flux
densities for Sgr A* (2 to 30 mJy), all of our near-infrared
measurements are consistent with a constant spectral index of
$\alpha$ = -0.6 $\pm$ 0.2 ($F_{\nu} \propto \nu^{\alpha}$).
Furthermore, this value is consistent with the spectral
indices observed at X-ray wavelengths during nearly all
outbursts; which is consistent with the synchrotron
self-Compton model for the production of the X-ray emission.
During the coordinated observations, one infrared outburst
occurs $\le$36 min after a possibly associated X-ray outburst,
while several similar infrared outbursts show no elevated
X-ray emission.  A variable X-ray to IR ratio and constant
infrared spectral index challenge the notion that the infrared
and X-ray emission are connected to the {\it same} electrons.
We, therefore, posit that the population of electrons
responsible for both the IR and X-ray emission are generated
by an acceleration mechanism that leaves the bulk of the
electron energy distribution responsible for the infrared
emission unchanged, but has a variable high-energy cutoff.
Occasionally a tail of electrons $\ga$ 1 GeV is generated, and
it is this high-energy tail that gives rise to the X-ray
outbursts. One possible explanation for this type of variation
is from the turbulence induced by a magnetorotational
instability, in which the outer scale length of the turbulence
varies and changes the high-energy cutoff.
\end{abstract}


\section{Introduction} %\label{intro}
Since its discovery, the compact radio source, \object{Sagittarius
  A*}, at the center of the Milky Way \citep{balick74}, now associated
with a 3.7 $\times$ 10$^{6}$ \msun\ supermassive black hole
\citep{schoedel03, ghez03, ghez05orbit}, has been noted for its
unusual emission properties.  Detections of emission from Sgr A* at
X-ray \citep{baganoff03} and infrared (IR; \citealt{genzel03flare,
  ghez04, ghez05lgs, eckart04}) wavelengths have offered new probes
into the central engine of our Galaxy.  One of the challenges for
models of this notably low luminosity source (10$^{36}$ erg s$^{-1}$
or, equivalently, 10$^{-9}$ L$_{Eddington}$) is to account for the
relationship, if any, between emission at different wavelengths.
Current models typically assume that all the emission originates
either in an accretion flow onto the supermassive black hole or in an
associated outflow \citep[e.g.,][]{falcke00,melia00,yuan02}. The radio
and IR photons are then presumed to be generated as synchrotron
emission from a highly energetic population of electrons and the X-ray
photons are produced by a combination of bremsstrahlung emission plus
either a continuation of the radio/IR synchrotron emission or a
synchrotron self-Compton component (SSC) of the radio/IR emission
\citep[e.g.,][]{baganoff01, markoff01, liu02, quataert02, yuan03}.

A new constraint on the origin of Sgr A*'s emission has come
from the observed variations in its emission properties as a
function of time.  At radio wavelengths, only modest emission
variations of a factor of two at most, and only occasionally
that much, have been detected on time scales from several
hours to months \citep[e.g.,][]{falcke99, bower02, zhao03,
herrnstein04, mauerhan05}. In contrast, dramatic variations
have been seen in the intensity of the X-ray emission, with
bursts that increase the X-ray emission by factors as large as
$\sim$160, that last roughly an hour, and that occur about
once a day \citep{baganoff01, baganoff02, goldwurm03,
porquet03}.  At near-infrared wavelengths, the emission varies
on many different time scales ranging from minutes to hours
\citep{genzel03flare, eckart04, eckart06, clenet04, clenet05,
eisen05, ghez04, ghez05lgs, gillessen06, krabbe06, zadeh06}.
For infrared outbursts lasting approximately an hour
(including both a rise and fall of the intensity), early 2
\micron\ studies estimate a flare frequency of 4 $\pm$ 2 times
per day for outbursts with peak intensities between 5--15 mJy
\citep{genzel03flare,eckart06} and more recent work at longer
wavelengths (3.8 \micron) indicate that peaks may occur up to
10 times per day \citep{hornsteinphd}.  Coordinated infrared
and X-ray observing campaigns have led to the detection of
five additional X-ray outbursts, all of which are roughly
coincident in time with infrared peaks \citep{eckart04,
eckart06, zadeh06}.  However, these outbursts show a wide
range of X-ray to IR peak ratios.  This could be attributed to
changes in the infrared spectral index \citep[e.g.,][]{yuan04,
liu06, zadeh06, bittner06}.  Recent measurements have
suggested variations of the near-infrared spectral index,
$-4\la\alpha\la1$, where $F_{\nu} \propto \nu^{\alpha}$, with
Sgr A* possibly becoming bluer when it is brighter
\citep{ghez05lgs, eisen05, gillessen06, krabbe06}.  However,
all but one of the IR spectral index measurements have so far
been made with spectroscopy, which is a challenging
measurement given the short wavelength bandpasses, low
emission levels, rapid time variations, and high local stellar
densities. Furthermore, none of the infrared spectral index
measurements have been made, so far, in coordination with
observations at other wavelengths.

An important first step in determining the physical cause of
these outbursts is to reliably identify the emission processes
responsible for creating the photons observed at various
wavelengths.  By measuring multiple properties of the
outbursts (e.g., intensity levels and spectral indices) at
multiple wavelengths (both during high and low emission
states), a clearer picture will emerge. For example, if the
X-ray emission is due to the same synchrotron process
invoked for the IR emission, then the spectral index from IR
to X-ray wavelengths should be the same (and thus the IR/X-ray
peak ratio would reveal the same spectral index). On the other
hand, an SSC model for the X-ray emission would require that
the typical spectral indices of the synchrotron (IR) and
SSC (X-ray) emission match each other, while the IR-to-X-ray
spectral index could vary.

In this paper, we present the first time-series measurements
of Sgr A*'s broadband infrared ($H-K^\prime$ and
$K^\prime-L^\prime$) colors, which were obtained with the
laser guide star adaptive optics system (LGS AO) on the
W. M. Keck II 10-meter telescope and most of which were taken
simultaneously with X-ray observations from the
\textit{Chandra X-Ray Observatory}.  By comparing to
spectroscopically identified stars in the Galactic Center,
these broadband colors can be converted into spectral indices
for Sgr A*. Additional natural guide star adaptive optics (NGS
AO) Keck observations of Sgr A*'s $L^\prime-M_s$ colors are
also reported.  The observations, data analysis, and results
are described in \S\ref{ir} and \S\ref{xray}, for the IR and
X-ray measurements, respectively.  These measurements show
that, while large changes in the intensity of the infrared
emission are detected, the infrared spectral index is constant
throughout all of our observations.  In \S\ref{discussion} we
discuss the implications these findings have on current
models. In particular, we argue that these observations
support the SSC model for the production of the 2--8 keV emission
and that a variable high-energy cutoff of the acceleration
process would be required to generate a variable IR/X-ray
intensity ratio while maintaining constant and equal IR and
X-ray outburst spectral indices.

\section{Keck Near-Infrared Adaptive Optics Data}\label{ir}

\subsection{Near-Infrared Observations}%\label{nirobs}
On 2005 July 31 and 2006 May 2 (UT), Galactic Center
observations in the $H$ ($\lambda_{o}$ = 1.63 \micron,
$\Delta\lambda$=0.30 \micron), $K^\prime$ ($\lambda_{o}$ =
2.12 \micron, $\Delta\lambda$ = 0.35 \micron), and $L^\prime$
($\lambda_{o}$ = 3.78 \micron, $\Delta\lambda$=0.70 \micron)
photometric bandpasses were conducted using the facility
near-infrared camera, NIRC2 (K. Matthews et al. 2006, in
preparation) behind the laser guide star AO system
\citep{wiz06, vandam06} on the W. M. Keck II 10-meter
telescope. With the narrow-field camera of NIRC2 (9.93 mas
pixel$^{-1}$), the 1024 $\times$ 1024 pixel InSb array
provides a field of view of 10\farcs2 $\times$ 10\farcs2.
\object{USNO-A2.0 0600-28577051} (R = 13.7 mag and
$\Delta$$r_{Sgr~A*} = 19\arcsec$) was used during the
observations to provide information on the tip/tilt term of
atmospheric aberrations, which can not be derived from the
laser guide star.  For more details regarding the general LGS
AO performance on the Galactic Center, see \citet{ghez05lgs}.
Measurements made on the night of 2005 July 31 with only the
tip/tilt loops closed on a bright star at zenith yielded a
seeing measurement of $\sim$0\farcs18 at 2 \micron\ (which
corresponds, theoretically, to 0\farcs23 at the Galactic
Center airmass of 1.52). Seeing conditions on the night of
2006 May 2 were much less favorable and conditions were
extremely variable throughout the night (in fact, due to the
much poorer atmospheric conditions, setup for the LGS AO
system was sufficiently time consuming that no official seeing
measurement was made). As such, only observations made at
airmass $\la$1.54 are usable for the 2006 May 2
observations. Resolutions (full widths at half maximum) and
Strehls for both epochs can be found in Table
\ref{nirobstable}.  In order to optimize observing efficiency
and eliminate any induced variability due to dithering, Sgr A*
was placed in an area of the detector free from bad pixels and
held fixed throughout the observations. Observations were made
by cycling through the $H$ (t$_{exp}$ = 7.4 s $\times$ 3
coadds), $K^\prime$ (t$_{exp}$ = 2.8 s $\times$ 10 coadds),
and $L^\prime$ (t$_{exp}$ = 0.5 s $\times$ 60 coadds) filters
repeatedly for 113 minutes in 2005 and 52 minutes in 2006,
with a three-filter cycle completed approximately every three
minutes. On 2005 July 31, these observations were interrupted
briefly (9 min) due to the requirement that the laser be
shuttered when its beam crosses the field of view of another
telescope on Mauna Kea.  Observations of a relatively dark
portion of the sky were conducted after the Galactic Center
observations to measure the background emission. In order to
reduce the effects of thermal emission from dust on the warm
image rotator mirror inside the AO enclosure, background
images at $L^\prime$ were obtained at two-degree increments
covering the full range of physical angles of the rotator
mirror during the Galactic Center observations.

Additional $K^\prime$ and $L^\prime$ NIRC2/LGS AO observations
of the Galactic Center were made on 2006 July 17
(UT). Integration times and coadds were the same as the 2005
July 31 epoch but during these observations, the telescope was
dithered semi-randomly within a box 0\farcs7 on a side
centered on Sgr A* with 3 pairs of $K^\prime$ and $L^\prime$
images taken at each dither position. Resolutions and Strehls
were comparable to the 2005 July 31 night and measurements on
a bright star at zenith yielded a seeing of 0\farcs35 at 2
\micron\ (see Table \ref{nirobstable}).  These observations
lasted approximately 187 minutes with a 16 minute outage in
the middle due to a technical problem with the LGS AO
system. Observations of the background sky/thermal emission
were accomplished in an identical manner to the observations
described above.

Complementary adaptive optics images of the Galactic Center
were obtained on 2005 July 16 (UT) using NIRC2 and the AO
system in its natural guide star mode \citep{wiz00a}. USNO-A2.0
0600-28577051 was again used but this time to provide
information on all terms of the atmospheric aberrations.
Observations in the $L^\prime$ and $M_s$ ($\lambda_{o}$ =
4.67 \micron, $\Delta\lambda$ = 0.24 \micron) photometric
bandpasses were interleaved by cycling through a 3 position,
0\farcs5 dither pattern on the array (to avoid dithering Sgr
A* into the noisy, bottom-left quadrant of the array), in
which two $L^\prime$ (t$_{exp}$ = 0.5 s $\times$ 120 coadds)
and three \Ms\ (t$_{exp}$ = 0.2 s $\times$ 600 coadds)
exposures were taken at each position.  While 0.2 s exposures
are technically possible using the full size of the array,
sub-arraying to 512 $\times$ 512 pixels greatly improves the
observing efficiency at \Ms\ and was used for all exposures on
this night.  Again, observations of the background sky/thermal
emission were taken at multiple rotator mirror angles after
the GC observations. Additionally, \Ms\ photometric standard
stars HD 1881 ($M_{s}$ = 7.17 $\pm$ 0.02), HD 3029
(7.70 $\pm$ 0.02), and SAO 3440 (7.04 $\pm$ 0.02) were observed in
order to calibrate the \Ms\ images \citep{leggett03}. Table
\ref{nirobstable} summarizes all of the Galactic Center IR
observations presented here.

\subsection{Near-Infrared Data Analysis}%\label{nirdata}
The standard image reduction steps of background subtraction,
flat-fielding, bad pixel repair, and optical distortion
correction were carried out on each exposure. As in
\citet{ghez05lgs}, the best background subtraction at longer
wavelengths ($L^\prime$ and \Ms), for both the LGS and NGS
data, was achieved by subtracting, from each exposure, the
average of three sky exposures that were taken at similar
rotator angles. Since the narrow-field camera of NIRC2
oversamples the point-spread function (PSF) at all wavelengths
(see Table \ref{nirobstable}), all images were binned by a
factor of two for computational efficiency. Additionally, due
to the lower signal-to-noise ratio (SNR) of the NGS AO data,
each \Ms\ image was made by averaging six individual exposures
while each $L^\prime$ image was made of four individual
exposures (two taken between, and one on either side of, the
corresponding two sets of \Ms\ exposures.)

Identification and characterization of point sources is
accomplished using StarFinder, an IDL PSF-fitting package
developed for astrometry and photometry in crowded stellar
fields \citep{diolaiti00}.  StarFinder is first run on an
average image made from all the data taken on each night as
well as on images made from three independent, equal-quality
subsets of the data for each bandpass. Sources having a range
of intensities are selected in StarFinder in order to
determine the PSF in the images.  The brightest stars, whose
cores are in the non-linear regime of the detector ($>$12,000
counts), are used to generate the PSF wings while fainter
sources are used to produce the PSF core.\footnote{IRS 7, IRS
16NE, IRS 16C, IRS 16NW, IRS 33E, IRS 33W, S1-23, and
+2.33+4.60 are used in all three LGS AO bandpasses.
Additionally, IRS 29N and IRS 16SW are used in the LGS AO $H$
and $K^\prime$ images while IRS 29NE1 is used in the LGS AO
$L^\prime$ images. Due to the sub-arraying used for the NGS AO
images, only IRS 16C, IRS 16NW, S1-23, and IRS 29NE1 remain in
the field of view for both $L^\prime$ and $M_{s}$ (star names
from \citealt{ghez98, genzel00}).}  For each bandpass, any
source that is detected in the average image (with a
correlation value of $\ge$0.8 in the 2005 July and 2006 July
data and $\ge$0.6 in the 2006 May and 2005 July data) as well
as at least two out of the three subsets (correlation $\ge$0.6
and $\ge$0.4 for the two groupings mentioned previously) is
considered a `validated' source.\footnote{Unfortunately, in
the lower quality 2006 May data set, while S0-17 is detected
in the average image (only 26 mas from Sgr A*), it is not
detected outright in any of the $H$ or \Lp\ subsets. However,
as it is a well known star from proper motion studies
\citep[e.g., ][]{ghez05orbit}, and this experiment relies on
accurately accounting for the emission contribution from all
nearby sources, it is deemed a valid source for the next step
of the analysis.} This list of sources is then used as the
input detection list for a modified version of StarFinder, run
on the individual images, that allows the intensity of the
sources to vary but holds their positions fixed.\footnote{If
the position of Sgr A* is left as a free parameter and allowed
to vary during its intensity variations, its photocenter is
very clearly biased towards the direction of the brightest
undetected emission when Sgr A* is at its lowest emission
levels. Since the direction of the bias depends on the
observed wavelength and is always in the direction of the next
nearest source of comparable emission, this displacement is
most certainly not physical.}

Images are photometrically calibrated using the apparent
magnitudes of several of the brightest stars in each
image. Specifically, we use measurements of IRS 16C ($H$ =
11.99, K = 9.83, $L^\prime$ = 8.14 mag) and IRS 16NW ($H$ =
12.04, K = 10.03, $L^\prime$ = 8.43 mag), from \citet{blum96}
at $H$ and $K^\prime$ and S.~A. Wright et al. (2007, in
preparation) at $L^\prime$, which have estimated uncertainties
of $\sim$4\%, 4\%, and 5\% at $H$, $K^\prime$, and $L^\prime$,
respectively.  Using the observed \Ms\ photometric standards,
we calibrate IRS 16C (\Ms\ = 8.08) and IRS 16NW (\Ms\ = 8.41)
in the average \Ms\ image for the night and then use them as
secondary calibrators for the individual images.  Absolute
calibration uncertainties at \Ms\ are estimated to be
$\sim$9\%. Determination of the uncertainties in the relative
photometry of Sgr A*, at all wavelengths, is carried out as in
\citet{ghez05lgs} by calculating the RMS flux density
variations between images for non-variable stars of comparable
magnitude. Dereddened flux densities are calculated assuming a
visual extinction of 29 $\pm$ 1 mag from S.~A. Wright et
al. (2007, in preparation) and an extinction law derived by
\citet{moneti01}, and then converted to flux densities with
zero points from \citet[][see footnotes in Table
\ref{nirobstable}]{tokunaga05}.

\begin{figure}
\plotone{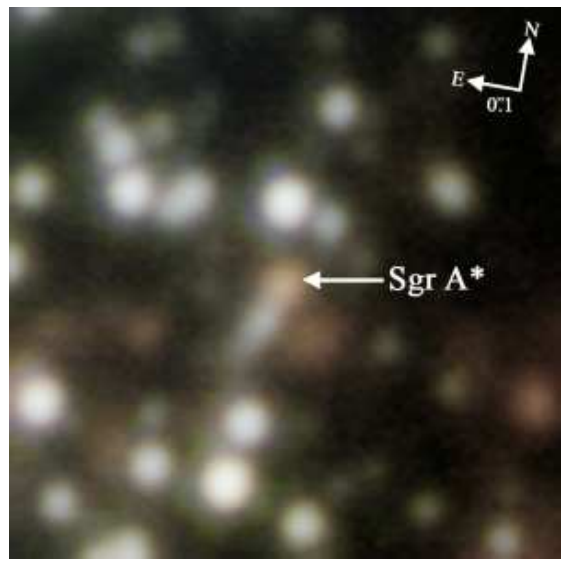} \figcaption[]{Central 1\arcsec\
$\times$ 1\arcsec\ region of the Galactic Center. This
false-color image is constructed from the average $H$ (blue),
$K^\prime$ (green) and $L^\prime$ (red) images taken on 2005
July 31 with the LGS AO system.  The high density of stars
seen in this image shows the need for a robust detection
algorithm in order to remove their contaminating emission,
which is much bluer than Sgr A*'s, and thereby to obtain an
accurate measurement of Sgr A*. The red emission immediately
to the SW of Sgr A* has been previously identified as a dust
feature projected along the line of sight toward Sgr A*
\citep{ghez05lgs}.\label{rgb}}
\end{figure}

The IR spectral indices are calculated from Sgr A*'s observed
colors and the relative interstellar reddening (color excess)
inferred from nearby stars. With this approach, the equation
for the spectral index, $\alpha$ (defined as $F_{\nu} \propto
\nu^{\alpha}$), is
\begin{equation}\label{spectraleq}
\alpha_{\lambda_{1}-\lambda_{2}} =\frac{-0.4
[(m_{\lambda_{1}}-m_{\lambda_{2}})_{obs} - 
E(\lambda_{1}-\lambda_{2})]
+
log(\frac{f_{0,\lambda_{1}}}{f_{0,\lambda_{2}}})}
{log(\frac{\lambda_{2}}{\lambda_{1}})},
\end{equation}
where $m_{\lambda_{n}}$ is the apparent magnitude of Sgr A* at
$\lambda_{n}$, E($\lambda_1-\lambda_2$) is the color excess
measured from nearby stars, and $f_{0,\lambda_{n}}$ is the
flux density of a 0 mag star at $\lambda_{n}$. Sgr A*'s colors
are derived from flux density estimates that have the same
time center. For the LGS AO data, this is achieved by
interpolating the flux densities at $K^\prime$-band as a
function of time to the flux density at the time of the
nearest measurement within the $H$- or $L^\prime$-band. In all
cases the difference between the time of the measurement and
the time to which the measurement is interpolated is always
less than 81 s. Since each pair of average $L^\prime$ and \Ms\
images in the NGS AO data set are centered on the same time,
no interpolation is necessary. By dereddening the apparent
colors of Sgr A* with the color excesses of nearby stars (see
Appendix), we avoid systematics associated with uncertainties
in the extinction law or photometric zero points.

\begin{figure}
\epsscale{1.0} 
\plotone{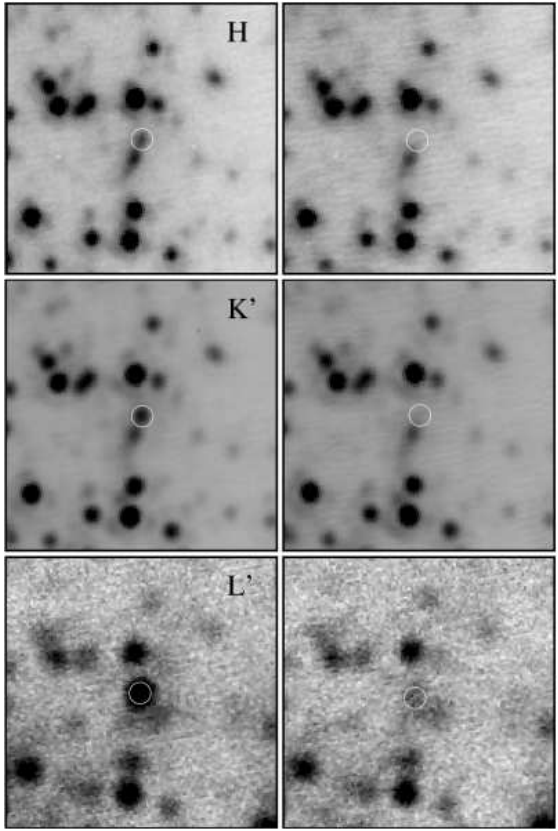} \figcaption[]{1\arcsec\
$\times$ 1\arcsec\ images showing Sgr A* (white circle) at its
highest (left) and lowest (right) states in $H$, $K^\prime$,
and $L^\prime$ on 2005 July 31 taken at 07:57 and 08:32 (UT),
respectively. The maximum and minimum recorded values are 9,
12, 15 mJy and 3, 2, 3 mJy at $H$, $K^\prime$, and $L^\prime$,
respectively. The StarFinder PSF-fitting routine consistently
detects a point source at the location of Sgr A* at all three
wavelengths with correlation values of $>$0.7.\label{minmax}}
 \end{figure}

\subsection{Near-Infrared Results}%\label{nirresults}
Figure \ref{rgb} shows a color composite of the central
1\arcsec\ $\times$ 1\arcsec\ $H$, $K^\prime$, \& $L^\prime$
images from 2005 July 31. This image shows the high density of
stars surrounding Sgr A*, necessitating that the flux from all
stars of comparable brightness to Sgr A* be accounted for.
This is especially important when the emission from Sgr A* is
at its lowest levels, as shown in the right panel of Figure
\ref{minmax}.  

\begin{figure}
\epsscale{1.0} \plotone{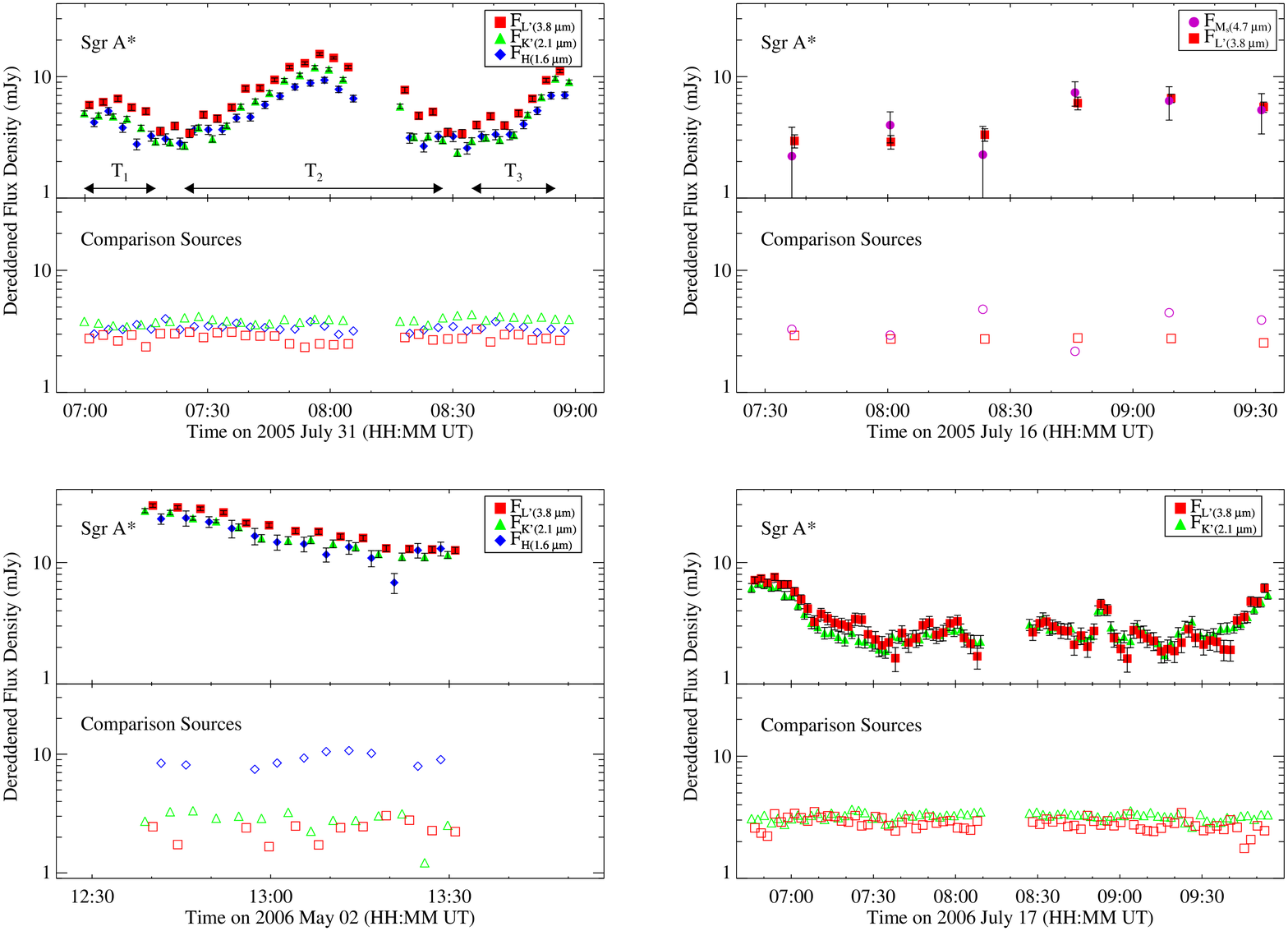} \figcaption[]{Light
curves for Sgr A* and nearby comparison stars of similar
brightness at each wavelength for 2005 July 31 (LGS AO; top
left), 2005 July 16 (NGS AO; top right), 2006 May 2 (LGS AO;
bottom left), and 2006 July 17 (LGS AO; bottom right). In all
but 2006 May 2, the comparison sources are chosen to match the
measurements of Sgr A* at its minimum values: S0-37 (ID8 in
\citealt{schoedel03}) at $H$ and $K^\prime$, S0-17 at
$L^\prime$, and S0-1 at $M_{s}$. (In the lower quality 2006
May 2 images, S0-37 was not detected in the $H$-band images
and has been replaced with S0-17.) Over four epochs, many
outbursts are detected, including three separate events during
the 2005 July 31 observations, marked as T$_1$, T$_2$, and
T$_3$.
\label{lightcurves}} 
\end{figure}

During our observations, Sgr A* exhibits a wide range of flux
densities as shown in Figure \ref{lightcurves}, which displays
the dereddened light curves of Sgr A* and comparison stars for
all of the 2005 and 2006 observations.  During our 2005 LGS AO
observations on July 31, Sgr A*'s light curve shows two clear
minima, with values of 2-3 mJy in all three filters,
separating emission events that can be associated with three
distinct peaks.  However, only one maximum is observed in its
entirety with peak levels of 9, 12, and 15 mJy in the $H$,
$K^\prime$, and $L^\prime$ filters, respectively.  During our
2005 July 16 observations, Sgr A* starts out in a low state of
3 and 2 mJy at $L^\prime$ and $M_{s}$, respectively, and rises
to 7 mJy in both filters by the middle of the observations.
While the 2006 May data set has the poorest image quality, it
is a particularly interesting observation because Sgr A* is
then detected in its highest IR outburst intensity measured to
date. At the beginning of the observations, Sgr A* is measured
at dereddened fluxes of 23, 27, and 30 mJy in the $H$,
$K^\prime$, and $L^\prime$ filters, respectively. In the
following 52 minutes, it falls to approximately 13 mJy in all
three filters. In the 2006 July LGS AO observations, Sgr A*'s
light curve shows a decaying flank from the outset, starting
out at 7 and 8 mJy in the $K^\prime$ and $L^\prime$ filters
and falling to 2-3 mJy in both filters over the course of 50
minutes. At the very end of these observations, the beginning
of yet another outburst event can be seen.

\begin{figure}
\epsscale{1.0} 
\plotone{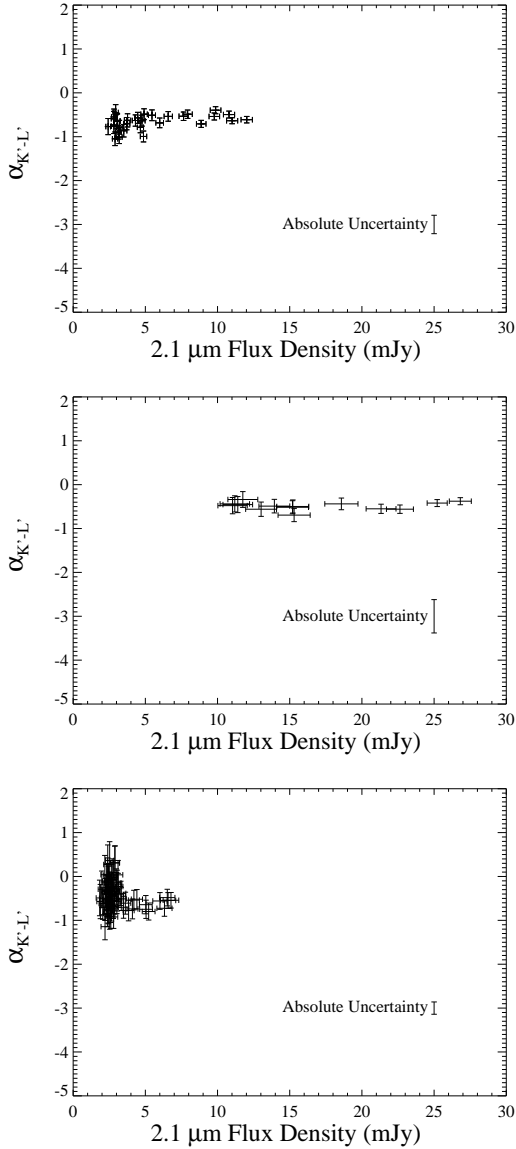}
\figcaption[]{Spectral index ($F_{\nu} \propto \nu^{\alpha}$)
vs. Sgr A* flux density derived from $K^\prime-L^\prime$ on
2005 July 31 (top), 2006 May 2 (middle), and 2006 July 17
(bottom). Within the measured uncertainties, Sgr A*'s spectral
index shows no variation with intensity over multiple outburst
events. The absolute uncertainty show in each plot is derived
from the uncertainty in the color excess correction (see
Appendix).
\label{alphavsflux}}
\end{figure}

Sgr A*'s spectral index appears to be independent of flux
density. This is most clearly seen in the spectral indices
derived from the dereddened $K^\prime-L^\prime$ colors, since
measurements from $H-K^\prime$ suffer from stellar background
contamination (see discussion below) and those from
$L^\prime-M_s$ have a much lower SNR due to the high thermal
background at \Ms. Figure \ref{alphavsflux} shows the
stability of $\alpha_{K^\prime-L^\prime}$ over flux density
variations spanning multiple outburst events. The plotted
uncertainties only incorporate those of the relative flux
density measurements of Sgr A*. Within the three runs in which
$K^\prime-L^\prime$ was measured, the spectral index is
constant with flux density. The larger spread of $\alpha$ at
flux densities below 4 mJy in 2006 July compared to 2005 July
is most likely a consequence of the poorer image quality,
which increases the effects of background contamination at low
flux levels.  For the run with the highest image quality, the
spectral index is consistent with a constant value to within
29\% over intensities ranging from 2 -- 12 mJy. Even with the
poorer seeing conditions in the 2006 May observations, which
span the largest change in flux densities (11 -- 27 mJy), the
spectral index is constant to within 19\%.

\begin{figure}
\epsscale{1}
\plottwo{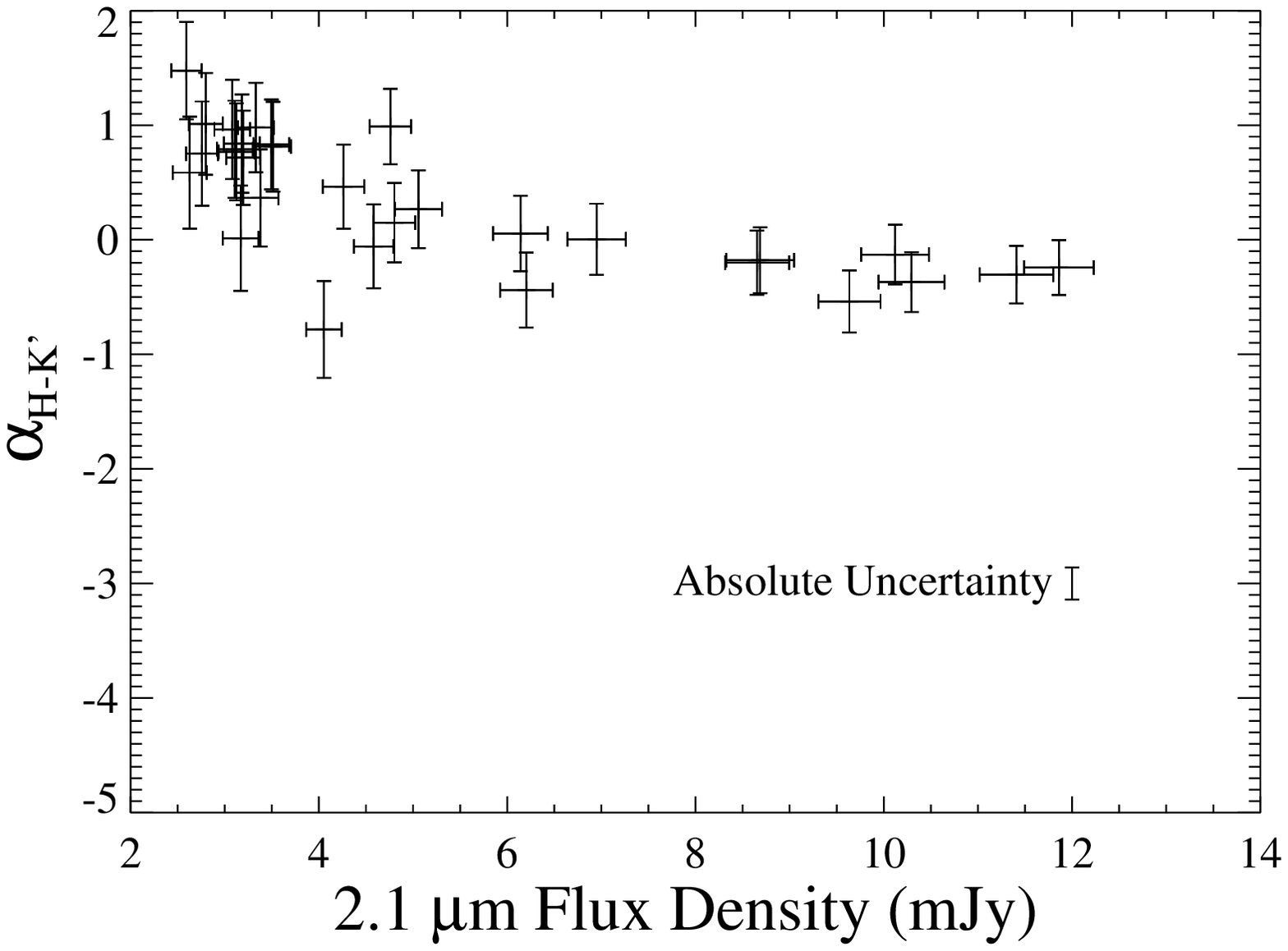}{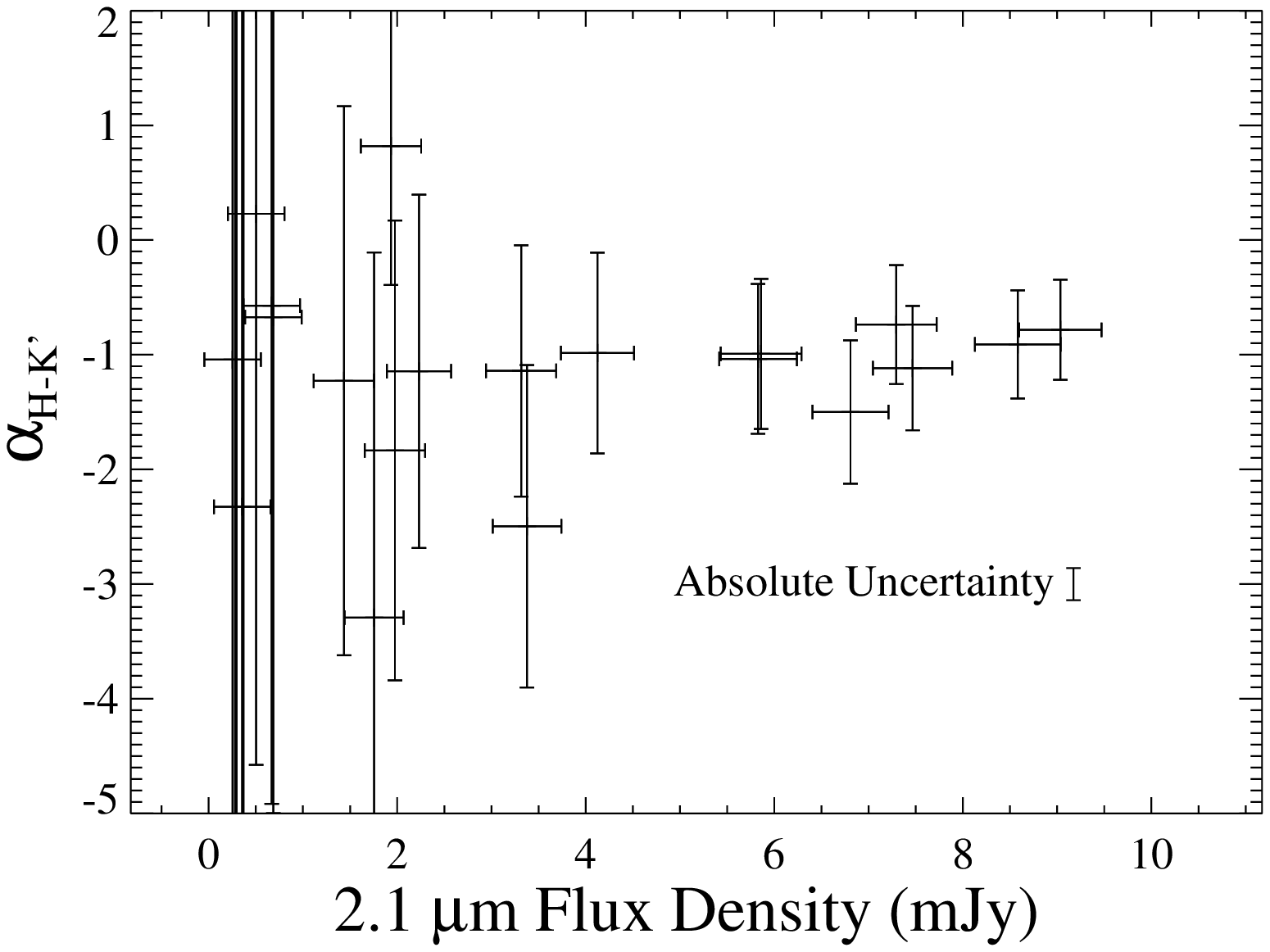}
\figcaption[]{(Left) Spectral index, $\alpha$ ($F_{\nu}
\propto \nu^{\alpha}$) vs. Sgr A* flux density derived from
$H-K^\prime$ on 2005 July 31 before the background
contamination has been removed.  While the higher flux
densities show the same stability trend as
$\alpha_{K^\prime-L^\prime}$, the spectral indices derived
below 4 mJy are much bluer, which we attribute to
contamination from the underlying stellar population. (Right)
The same spectral indices after the background contamination
is removed by subtracting the average level of emission during
the first minimum ($H$=3.2 $\pm$ 0.2, $K^\prime$=2.8 $\pm$
0.1, $L^\prime$=3.6 $\pm$ 0.2 mJy). While the errors have
increased due to the subtraction technique, these spectral
indices are consistent with the spectral indices at other
wavelengths.\label{alphavsflux_hk}}
\end{figure}

Sgr A*'s spectral index is also constant over wavelengths
ranging from 1.6 -- 4.7 microns. At shorter wavelengths, the
spectral indices derived at flux densities below 4 mJy show a
trend blueward with decreasing intensity (See Figure
\ref{alphavsflux_hk}, left). Since this effect is not seen at
the longer wavelengths, we attribute this effect to
contamination from the underlying stellar population (which is
much bluer than Sgr A*; see Figure \ref{rgb}). To investigate
this in more detail, this local background contamination is
removed from the $H-K^\prime$ measurements by subtracting off
the minimum level observed at the position of Sgr A* during
the observations, which is taken to be the average level
detected during a 10 minute period in the first observed
minimum beginning at $\sim$07:20 for the 2005 July data set
and the average of the last three points for the 2006 May data
set.  With this approach, the 2005 $H-K^\prime$ spectral
indices are not only constant with respect to outburst
intensity (see Figure \ref{alphavsflux_hk}, right), but also
consistent with the results from longer wavelengths shown in
Figure \ref{alphavsflux} and Table \ref{indices}.  Applying
this subtraction technique to the longer wavelength
measurements in 2005 July ($K^\prime-L^\prime$ and
$L^\prime-M_s$) results in no significant change in the
spectral indices; thus supporting the hypothesis that
the contaminating background is from the blue stellar
population. A similar trend is seen in the 2006 May
$H-K^\prime$ and $K^\prime-L^\prime$ measurements. We,
therefore, correct only the $H-K^\prime$ spectral indices for
local background contamination.

Table \ref{indices} presents the weighted average of $\alpha$
derived from various wavelength pairs over multiple observing
runs, including a recalculation of the spectral index for one
pair of $K^\prime$/$L^\prime$ observations in 2004
(\citealt{ghez05lgs}).\footnote{The original calculation used
an assumed extinction law to derive $\alpha$ while our
recalculation uses the method outlined above. See discussion
in Appendix.} The reported uncertainties include the
propagated uncertainties of the observed color of Sgr A* as
well as the uncertainties from the dereddening method.  All of
these measurements are consistent with each other within their
uncertainties and, when averaged together, imply a spectral
index for Sgr A* during these observations of -0.63 $\pm$ 0.16
(where we have taken the weighted average of the three
different color pairs from Table \ref{indices} after
quadratically summing the two sources of error). This spectral
index appears to be independent of intensity (F$_{IR}$ = 2 --
30mJy), time (multiple outbursts over three years), or
wavelength (1.6 -- 4.7 microns). 

Several comments can be made about these results when compared to
previous results in the literature. While photometrically derived
spectral indices of Sgr A* are obtained in an inherently different
manner than those obtained directly from spectra, they agree within
approximately two sigma with OSIRIS K-band (2.02--2.38 \micron)
spectra obtained at a comparable emission level (i.e., $\alpha$ = -2.6
$\pm$ 0.9 when F$_{Sgr~A*}$ = 6.1 mJy; \citealt{krabbe06}).  Other
results from SINFONI K-band spectra (1.95--2.45 \micron) which suggest
a correlation of spectral index with emission intensity
\citep{gillessen06} are not as discrepant as they may seem. Of the
three methods presented by the \citeauthor{gillessen06}, the `off
state subtraction' method, which most closely resembles the method
presented here, results in a data set that is statistically consistent
with no spectral index/intensity correlation down to 2 mJy, and that
method gives $\alpha_{bright~ state} = -0.6 \pm 0.2$, which is in
excellent agreement with the value that we derive above. The apparent
correlation seen from the other two methods may be explained by the
local background contamination that becomes significant at flux
densities below 4 mJy.  A local aperture background at K-band could
sample the contamination from the surrounding population of blue stars
and the subtraction of an overly blue background would lead to a
redder appearance for Sgr A* at low levels, where the background
emission becomes comparable to Sgr A*'s intrinsic emission.
Furthermore, $\alpha_{preflare}$ from \citeauthor{gillessen06}, as
well as the spectral indices presented by \citet{eisen05}, reporting
extremely red spectral indices for Sgr A* ($\alpha\la-3$), were made
at very low flux levels (F $\la$ 2 mJy), where the complications with
the background make this a very challenging measurement to make.
Therefore, we suggest that the apparent variation in the IR color of
Sgr A* found in previous experiments could arise as a consequence of
background contamination in this extremely crowded field.

\section{Chandra X-ray Data}\label{xray}
\subsection{ACIS X-ray Observations and Data Analysis}%\label{xrayobs}
X-ray observations of the Galactic Center were conducted with
the \textit{Chandra X-Ray Observatory} \citep{weisskopf96}
using the imaging array of the Advanced CCD Imaging
Spectrometer (ACIS-I; \citealt{garmire03}) from 2005 July 30
19:44 -- 2005 July 31 09:10 and 2006 July 17 04:17 -- 2006
July 17 12:37 (UT). The last two hours of the 2005
observations and three hours of the 2006 observations
overlapped with the Keck LGS AO observations from the ground.
The data were acquired and reduced as described by
\citet{baganoff01,baganoff03} \citep[see also][]{eckart04,
eckart06}.  Briefly, the instrument was operated in timed
exposure mode with detectors I0-3 and S2 turned on. The time
between CCD frames was 3.141 s and the event data were
telemetered in faint format.  The data were analyzed using the
Chandra Interactive Analysis of Observations (CIAO) v3.2
software package\footnote{http://cxc.harvard.edu/ciao/} with
CALDB v3.1.0. Source counts in the 2--8 keV band were
obtained from aperture photometry on Sgr A* using an aperture
radius of 1\farcs5 and a surrounding sky annulus extending
from 2\arcsec\ to 4\arcsec, excluding regions around discrete
sources and bright structures.

\subsection{X-ray Results}%\label{xrayresults}
\begin{figure}
\epsscale{1.0} \plotone{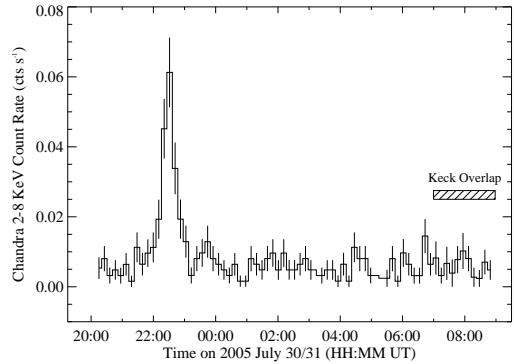}
\figcaption[]{Chandra 2-8 keV light curve of 2005 July 30/31
including the overlap with ground-based Keck observations. The
\textit{Chandra} observations begin on 2005 July 30 20:15:12.0
and end on 2005 July 31 08:49:04.0 (UT). While one significant
outburst is detected, peaking at $\sim$22:30 on July 30, it is
not simultaneous with the Keck IR
observations.\label{xraylightcurve}}
\end{figure}

\begin{figure}
\epsscale{1.0} \plotone{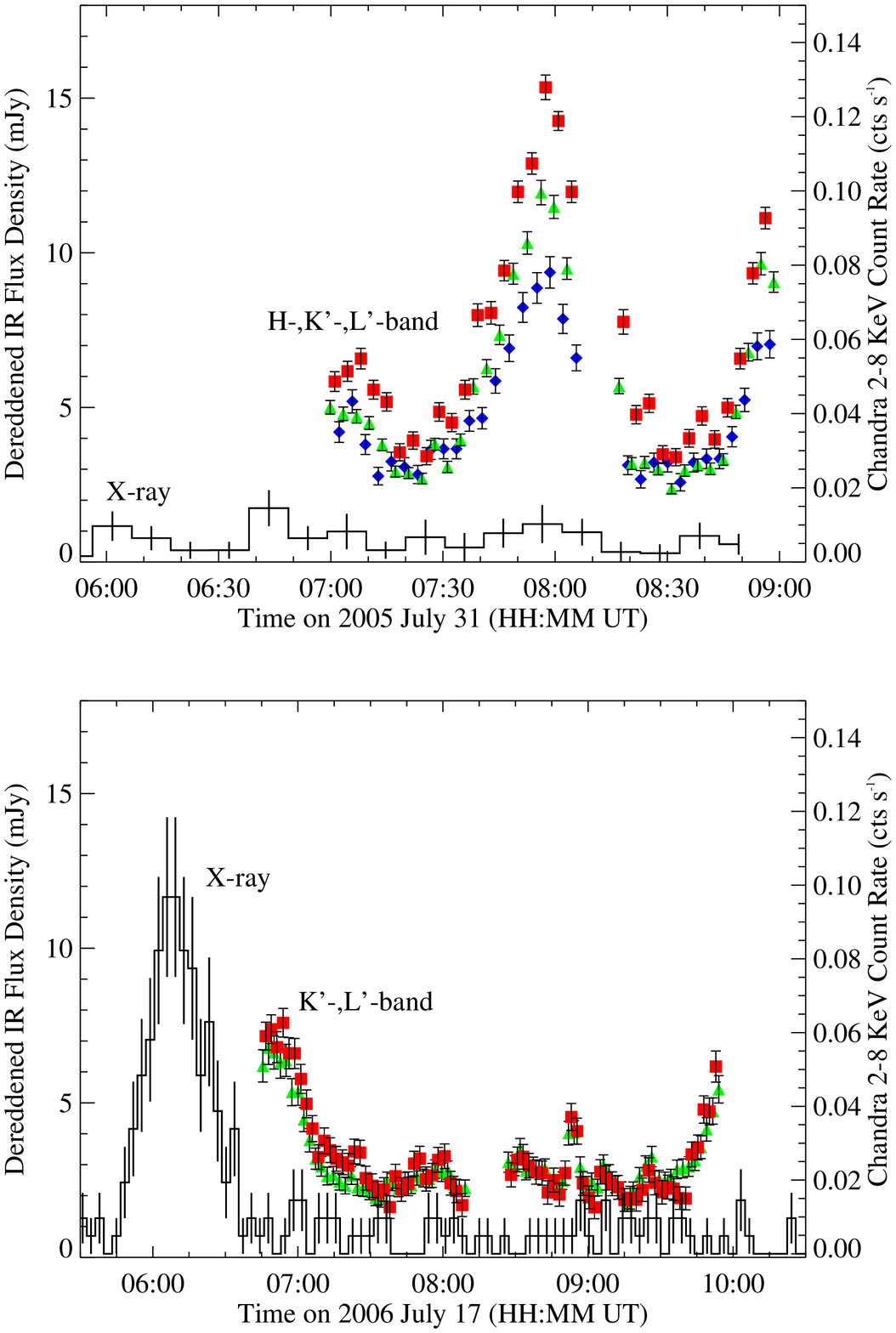}
\figcaption[]{The overlapping 2005 (top) and 2006
(bottom) IR (left axis) and X-ray (right axis) light curves
(see Figure \ref{lightcurves} for a description of the IR
symbols). No significant X-ray variations occurred within
$\sim$8.5 hours of the start of the 2005 IR observations. On
the other hand, $\sim$36 minutes before the beginning of the
2006 IR observations, \textit{Chandra} detected a factor of 20
outburst. As previous coordinated campaigns have detected IR
variations persisting more than 36 minutes after the
associated X-ray emission dropped below the constant level of
the diffuse emission from within the central parsec
\citep{eckart04,eckart06}, we associate the IR decay at the
beginning of our IR observations with the X-ray outburst.
\label{chandrakeck}}
\end{figure}

Figure \ref{xraylightcurve} shows the 2005 X-ray light curve
marked with the period of overlapping Keck IR data. Figure
\ref{chandrakeck} (top) shows the 2005 Keck light curves in
the $H$-, $K^\prime$-, and $L^\prime$-bands along with the
simultaneous part of the X-ray light curve.  While a
factor-of-twenty X-ray outburst is detected, it is not
simultaneous with the Keck IR observations. Here we restrict
our analysis to the period overlapping the Keck observations.
A Bayesian blocks analysis, as described in \citet{eckart04},
is performed on the X-ray light curve without background
subtraction.  This analysis indicates that Sgr A*'s X-ray
light curve is consistent with no variability at the 90\%
confidence level during the entire remaining period after the
main outburst.  The mean count rate of the
background-subtracted light curve from $\sim$23:00 - 07:00 is
4.29 $\pm$ 0.68 cts ks$^{-1}$ corresponding to a 2--8 keV
luminosity of 1.8 $\pm$ 0.3 $\times$ 10$^{33}$ erg s$^{-1}$.
During the Keck overlap, the mean count rate is 4.94 $\pm$
1.54 cts ks$^{-1}$, which lies within the 90\% confidence
interval, computed according to \citet{gehrels86}, for the
period cited above.  Therefore, we adopt this 90\% confidence
interval as an upper limit during the Keck observations,
corresponding to 2.1 $\times$ 10$^{33}$ erg s$^{-1}$.  Thus,
the X-ray luminosity does not change by more than $\sim$16\%
during the 2005 IR outburst observed with Keck.

In 2006, a strong ($\sim$20$\times$) X-ray outburst is
detected and it peaks at 4.0 $\times$ 10$^{34}$ erg s$^{-1}$
about 36 minutes before the Keck observations begin (Figure
\ref{chandrakeck}, bottom).  While there is a non-negligible
chance that the Keck IR outburst is unrelated to this X-ray
event (10\% given IR outbursts occurring four times a day,
25\% for outbursts ten times per day) previous coordinated
activity at IR and X-ray wavelengths have shown that the IR
decay can persist 36 minutes (and longer) after the presumably
associated X-ray peak (\citealt{eckart04}, Figure 4;
\citealt{eckart06}, Figure 9). Furthermore, all X-ray
outbursts which have had simultaneous IR observations, have
shown apparent associated IR outbursts. The probability of a
second IR event occuring within 36 minutes of the X-ray peak
(assuming that a unseen IR peak occurred simultaneously with
the X-ray peak and our IR observations captured a second,
unrelated outburst) is only 1\%(6\%) for outbursts occuring
four (ten) times per day.  This strengthens the possible
association here. We note that the decreasing X-ray emission
cannot be tracked for as long as the IR emission because of
\textit{Chandra's} lower angular resolution combined with the
bright, steady diffuse X-ray emission in the central
parsec. Additionally, the flatness at the beginning of the IR
light curve resembles short time-scale non-monotonic
fluctuations that are seen in the structure of these other
coordinated outbursts, and does not necessarily imply that a
peak has occurred at the beginning of the IR
observations. Thus, our data are consistent with no
temporal displacement between the X-ray and IR peaks.  It is,
therefore, possible that the elevated IR state observed in our
2006 July observations is associated with the X-ray activity
that immediately precedes it.

\section{Discussion} \label{discussion}
Our observations show that Sgr A*'s infrared spectral index is
constant, with a value of $\alpha$ = -0.6 $\pm$ 0.2 ($F_{\nu}
\propto \nu^{\alpha}$).  The lack of dependence upon intensity
is quite notable, given the wide range of infrared intensities
measured during our experiment (2--30 mJy).  While our low
points are comparable to some of the lowest measurements ever
made of Sgr A*, the peak emission detected in this study is
the highest infrared measurement recorded thus far
\citep{genzel03flare, eckart04, eckart06, clenet04, clenet05,
eisen05, ghez04, ghez05lgs, gillessen06, krabbe06,
zadeh06}\footnote{Due to the different reddening laws used by
various authors, all previous measurements are scaled by
comparing the value of the comparison source reported in each
study with the same source from this study. When no comparison
source is given, the stated reddening law is used.}. The IR
emission is most likely from synchrotron photons, as
demonstrated by the recent reports of polarized IR emission
from Sgr A* \citep{eckart06b, meyer06a, meyer06b}. Our
measured value of the spectral index of -0.6 indicates that
the synchrotron emission is optically thin and arises from an
electron energy distribution with
$N(\gamma_e)\propto\gamma_e^{-p}$, where
$p$~=~1-2$\alpha$~=~2.2 and $\gamma_e$ is the Lorentz factor
of the electrons. In order to associate particular electron
energies, or $\gamma_e$, with the observed IR synchrotron
emission, knowledge of the magnetic field strength is required
($\nu_{sync}=\frac{3}{4\pi}\gamma_e^2\frac{eB}{m_ec}$ and
where E$_e = \gamma_e m_ec^2$.)  Although highly uncertain, if
we adopt a magnetic field strength of B$\sim$20 G, as has been
suggested by others \citep[e.g., ][]{markoff01, yuan03}, then
$\sim$ GeV electrons ($\gamma_e\approx$2000) are responsible
for producing the observed IR emission. It has been generally
assumed that the intensity variations are caused by
acceleration events, possibly caused by an enhancement of the
mass accretion rate or magnetic reconnection events within the
accreting gas, which enable electrons to reach such high
energies \citep[e.g.,][]{markoff01, liu02, yuan03,yuan04,liu06,tagger06}.  
In order to produce IR intensity changes without
changing the IR spectral index, the electron acceleration
mechanism has to change only the total number of electrons in
this energy range while leaving their overall energy
distribution unchanged.

If the X-ray emission arises from inverse Compton scattering of
synchrotron photons, then the X-ray spectral index should match that
of the synchrotron emission. Examining the X-ray spectral indices from
all previously reported outbursts, we find that the X-ray spectral
indices\footnote{In the literature, most authors present the X-ray
  photon index ($\Gamma$), rather than the spectral index
  ($\alpha$). In the convention used in this paper, $\alpha =
  1-\Gamma$ and all photon indices obtained from the literature have
  been converted to spectral indices for comparison.} are consistent
(within two sigma) with the infrared spectral index derived here, in
all but the most extreme X-ray outburst, and have
$\langle\alpha_{X-ray}\rangle = -0.3 \pm 0.6$.\footnote{The
  uncertainty here comes from the RMS of all reported values, and
  given the large uncertainty on many of the measurements, this
  dispersion in values is not necessarily indicative of variation;
  $\alpha_{X-ray}$ = 0.0$_{-0.7}^{+0.8}$ \citep{baganoff01},
  -0.3$_{-0.4}^{+0.5}$ \citep{baganoff02}, 0.1 $\pm$ 0.5
  \citep{goldwurm03}, -1.5 $\pm$ 0.3 \citep{porquet03}, -0.5 $\pm$
  0.5, -0.9 $\pm$ 0.5 \citep{belanger05}, 0.5$_{-1.3}^{+0.9}$
  (F.~K. Baganoff et al. 2007, in preparation, 0.0$_{-1.6}^{+1.0}$
  (D.~P. Marrone et al. 2007, in preparation).}  As the largest X-ray
outburst dwarfs all other known X-ray outbursts by more than a factor
of three \citep{porquet03}, this event may represent a very different
scenario than is producing the more common X-ray events.
Nevertheless, the similarity of the typical X-ray and IR spectral
indices is completely consistent with the SSC model for the production
of the X-ray photons.

In contrast to the uniformity of the infrared and X-ray spectral
indices, the ratio of infrared to X-ray intensity appears to be quite
variable.  During our 2006 coordinated observations, decaying IR
emission is seen in association with a significant X-ray outburst. In
fact, all previous IR observations obtained during significant X-ray
intensity variations have shown similar associations \citep{eckart04,
  eckart06, zadeh06}. On the other hand, no elevated X-ray emission is
seen during our 2005 coordinated observations, in spite of the large
IR outburst that occurred then.  Similar occurrences of IR peaks with
no X-ray outbursts are reported by \citet{zadeh06}.  Previously, the
IR and X-ray photons have been theoretically linked through either a
synchrotron or SSC mechanism and it has been shown that a variation
in the ratio of X-ray to IR peak intensities would be accompanied by a
change in the infrared spectral index or, equivalently, in the slope
of the electron energy distribution \citep[e.g.,][]{yuan04,liu06b,
  bittner06}. \citet{zadeh06} suggest that a variation in the magnetic
field strength could be responsible for the variable IR/X-ray
intensity ratio observed in the double peak structure seen during
their coordinated observations (see \citealt{zadeh06}, Figure 9).  If
the magnetic field is reduced in an event, for example, then there are
two possible ways to account for the observed constancy of the
spectral index. If the IR synchrotron photons Compton scatter off the
$\sim$25 MeV electrons responsible for the submillimeter emission then
a reduction in the magnetic field in the IR emitting region would
result in a reduction of the infrared synchrotron emission and a
proportional reduction of the X-ray SSC emission (and, thus, would
display a constant IR/X-ray intensity ratio). If, on the other hand,
the submillimeter photons are Compton scattered by the same $\sim$1
GeV electrons in the hot spot that produce the IR synchrotron
emission, then a reduction in the magnetic field would result in the
IR synchrotron only without changing the X-ray SSC emission and the
IR/X-ray ratio would decrease.  However, both of these are
inconsistent with the observations.

The variation in the ratio of IR and X-ray peak intensities, despite a
constant spectral index at both IR and X-ray wavelengths, suggests
that one may not be able to connect the infrared and X-ray emission to
the \textit{same} electrons. Instead, we posit that while the
acceleration mechanism leaves the bulk of the electron energy
distribution unchanged, the upper energy cutoff ($\gamma_{max}$) is
not constant over every event (a variation of which was explored by
\citealt{yuan04}).  In this scenario, the X-ray photons are produced
by electrons that have higher energies (e.g. E$\ga$1 GeV for B$\sim$20
G) than those that give rise to the IR synchrotron photons. These
electrons can give rise to the X-ray emission in two ways. In one
scenario, the electrons produce synchrotron emission that is
upscattered off of the population of $\la$25 MeV electrons responsible
for the submillimeter emission. (These same low-energy electrons would
scatter the near-IR photons in the observable IR bands to energies
below 2 keV, which is undetectable by \textit{Chandra} due to
interstellar absorption.)  In an alternative scenario, $\ga$1 GeV
electrons act as the scattering population for the submillimeter
synchrotron photon field generated by the $\la$25 MeV electrons and
produce a similar level of observed X-ray activity \citep{zadeh06}.
In either case, the variation in X-ray emission is caused by the fact
that these $\ga$1 GeV electrons may not always be produced in the
acceleration event if $\gamma_{max}$ is not high enough. A variable
$\gamma_{max}$ could be the result of a variation of the largest
turbulence scale sizes accompanying different instances of a disk
instability.  Therefore, it is this occasional high-energy population
of electrons that gives rise to the relatively infrequent X-ray
outbursts. Furthermore, as energies below $\gamma_{max}$ are always
created during the acceleration event, X-ray outbursts, when present,
would always show correlated infrared activity (as is the case with
all current NIR observations during X-ray outbursts;
\citealt{eckart04,eckart06,zadeh06}). While this simple model can
explain a constant IR/X-ray intensity ratio when conditions are
suitable for creating the X-ray photons, physical parameters such as
the magnetic field, source size, and electron density in the emitting
region \textit{could} easily change from outburst to outburst,
resulting in a change in the IR/X-ray intensity ratio during
coordinated outbursts (for example, the IR/X-ray spectral index
between the two peaks seen by \citealt{zadeh06} changes by only
$\sim$0.14). Additionally, as suggested by \citet{yuan04}, a variable
$\gamma_{max}$ could explain the anomalous, soft X-ray spectral index
detected by \textit{XMM-Newton} \citep{porquet03}.

\section{Summary} %\label{summary}
The broadband infrared imaging method presented here, when
combined with PSF fitting photometry, allows for robust
measurements of the color of Sgr A* during an outburst due to
the short, high quality images that can be taken throughout
the event.  By comparing with several spectroscopically
identified stars, we are able to present spectral indices that
are not dominated by the uncertainties in the interstellar
reddening law. We find the most reliable spectral index
measurements are obtained from the $K^\prime-L^\prime$ colors,
due to the challenges associated with accounting for the blue
stellar background when Sgr A* is its lowest emission states
in the $H-K^\prime$ color; this challenge may also explain the
discrepant spectral indices in the literature.

From observations taken between 2004--2006 we find a spectral
index of $\alpha$ = -0.63 $\pm$ 0.16 ($F_{\nu} \propto
\nu^{\alpha}$) that is independent of infrared intensity or
wavelength throughout a variety of emission levels, including
the largest recorded IR outburst from Sgr A* to date (from 1.6
to 3.8 \micron).  Additionally, our coordinated observations
show that this spectral index is constant regardless of X-ray
emission activity (or lack thereof). A review of the spectral
indices reported for all reported X-ray outbursts reveals a
remarkable agreement with the IR spectral index presented
here, thus bolstering the case that the X-ray photons are the
result of synchrotron self-Compton emission.  Due to the
apparent stability and similarity of the infrared and X-ray
spectral indices despite wide ranges of reported infrared to
X-ray flux ratios, we suggest that X-ray photons originate
from a high-energy population of electrons occasionally
created by the same mechanism responsible for accelerating the
infrared emitting electron population.  Future monitoring of
Sgr A*'s IR spectral index in coordination with X-ray
observations are necessary to confirm that the infrared
spectral index remains constant throughout the peak of an
X-ray outburst.  In addition, the detection of an X-ray
outburst without a corresponding IR outburst would present a
serious challenge to the model presented here.

\acknowledgments 
The authors thank Randy Campbell, Al Conrad,
Steven McGee, Madeline Reed and the entire Keck LGS AO team
for their help in obtaining these observations, Emiliano
Diolaiti for help in modifying the StarFinder code, and Siming Liu
for useful discussions.  Support for this work was provided by
NSF grant AST 04-06816 and the NSF Science and Technology
Center for Adaptive Optics, managed by UCSC (AST
98-76783). SDH was also supported by a Dissertation Year
Fellowship provided by the University of California. FKB was
supported by NASA through Chandra award G05-6093X. The
infrared data presented herein were obtained at the W. M. Keck
Observatory, which is operated as a scientific partnership
among the California Institute of Technology, the University
of California and the National Aeronautics and Space
Administration. The Observatory was made possible by the
generous financial support of the W. M. Keck Foundation. The
authors wish to recognize and acknowledge the very significant
cultural role that the summit of Mauna Kea has always had
within the indigenous Hawaiian community.  We are most
fortunate to have the opportunity to conduct observations from
this mountain.

\textit{Facilities:} \facility{Keck:II (NIRC2)}, \facility{CXO (ACIS-I)}

\begin{appendix}
\section{Color Excesses From Nearby Stars}%\label{colorexcess}
Before the true spectral index of Sgr A* can be calculated,
the observed magnitudes must be appropriately dereddened. Due
to the visual extinction of A$_v\approx$29 towards Sgr A*, the
actual extinction values used at infrared wavelengths can have
a dramatic effect on the colors obtained for Sgr A*. Previous
attempts to calculate the spectral index of Sgr A* have used
several different methods for this dereddening making direct
comparisons difficult. In their imaging study of Sgr A*,
\citet{ghez05lgs} use an A$_v$ = 30 and an extinction law from
\citet{moneti01} to correct for extinction, while
\citet{krabbe06} and \citet{gillessen06} divide their spectral
measurements of Sgr A* by the spectrum of S0-2 and assume a
spectral type of S0-2 from previous studies
\citep{ghez03,eisen05}. As there exists a large discrepancy in
the literature for the extinction law towards the Galactic
Center as well as the actual value for A$_v$ \citep[see,
e.g.,][]{clenet01, scoville03, viehmann05}, correcting stars
with known spectral-types provides the least uncertainty. This
method also eliminates any potential errors caused by
photometric zero point calibrations.

We use several stars in the central 0\farcs5 with
well-determined spectral types \citep{ghez03, eisen05} to
derive an average color excess with which to deredden our
apparent colors of Sgr A*. Due to stellar crowding and the
larger PSF size and lower SNR in some of the longer wavelength
images, not all sources are detected in all images.  Table
\ref{compstars} lists the color excesses obtained for each
star as well as the weighted average used for dereddening. As
the spread in the color excesses calculated for various stars
around Sgr A* shows the uncertainties from calculating the
true color excess appropriate for the extinction towards the
Galactic Center, we take the RMS of the color excesses as the
associated error from this method. This value is then used to
obtain the absolute uncertainties on the spectral index
derived for Sgr A* for each wavelength pair in each observing
run (an error in the color excess propagates as an error in
the spectral index as $\sigma_{\alpha_{\lambda_1-\lambda_2}} =
\frac{\sigma_{E(\lambda_1-\lambda_2)}}{2.5log_{10}(\lambda_2/\lambda_1)}$).

An additional absolute uncertainty comes from the
uncertainties in the colors of the spectral types of B stars,
as reported by \citet[$\sigma_{H-K}=0.08$,
$\sigma_{K-L}=0.09$, and $\sigma_{L-M}=0.13$;][]{ducati01}.
These uncertainties in colors result in uncertainties of 0.27,
0.14, and 0.57 for $\alpha_{H-K^\prime}$,
$\alpha_{K^\prime-L^\prime}$, and $\alpha_{L^\prime-M_{s}}$,
respectively. However, as they are systematic errors that do
not change between observing runs within a given wavelength
pair, they only need to be taken into account when comparing
across multiple wavelengths. As such, they are not included in
the absolute uncertainties plotted in Figures
\ref{alphavsflux} and \ref{alphavsflux_hk}, but are included
in the values reported in Table \ref{indices} and the final
value of $\alpha$ = -0.63 $\pm$ 0.16.
\end{appendix}

\clearpage
\begin{landscape}
\input{tab1.tex}

\clearpage
\end{landscape}
\input{tab2.tex}
\input{tab3.tex}

%\clearpage
%\thispagestyle{empty}
\end{document}

%% file: tab1.tex
\begin{deluxetable}{cccccccccccc}
\tabletypesize{\tiny}
\tablecaption{Summary of Infrared Observations of Sgr A*}
\tablewidth{0pt}
\tablehead{
\colhead{Epoch} &
\colhead{AO Mode} &
\colhead{Coordinated} &
\colhead{Bandpass} &
\colhead{Start Time} &
\colhead{End Time} &
\colhead{N$_{obs}$} &
\colhead{T$_{exp,i}$} &
\colhead{Strehl} &
\colhead{FWHM} &
\multicolumn{2}{c}{Peak Emission\tablenotemark{a}} \\
\colhead{(UT)} &
\colhead{} &
\colhead{with \textit{Chandra}} &
\colhead{} &
\colhead{(UTC)} &
\colhead{(UTC)} &
\colhead{} &
\colhead{(s)} &
\colhead{} &
\colhead{(mas) }&
\colhead{Observed (mag)}& 
\colhead{Dereddened (mJy)\tablenotemark{b}} 
}
\startdata
2005 July 31 & LGS & Yes & \Hp~ & 07:02:14 & 08:57:25 & 31 &  22 & 0.18 $\pm$ 0.03 & 63 $\pm$ 2 & 17.76 $\pm$ 0.06 &  9.4 $\pm$ 0.5\\
             &     &     & \Kp  & 06:59:50 & 08:58:28 & 31 &  28 & 0.34 $\pm$ 0.02 & 62 $\pm$ 2 & 15.11 $\pm$ 0.04 & 11.6 $\pm$ 0.4\\
             &     &     & \Lp  & 07:01:03 & 08:56:14 & 31 &  30 & 0.70 $\pm$ 0.06 & 80 $\pm$ 1 & 12.09 $\pm$ 0.03 & 15.4 $\pm$ 0.4\\
2006 May 02  &     & No  & \Hp~ & 12:41:35 & 13:28:35 & 13 &  22 & 0.02 $\pm$ 0.01 &110 $\pm$ 9 & 16.79 $\pm$ 0.11 & 22.9 $\pm$ 2.3\\
             &     &     & \Kp  & 12:38:53 & 13:29:44 & 14 &  28 & 0.07 $\pm$ 0.02 &104 $\pm$10 & 14.23 $\pm$ 0.04 & 26.8 $\pm$ 1.1\\
             &     &     & \Lp  & 12:40:12 & 13:31:01 & 14 &  30 & 0.38 $\pm$ 0.03 &103 $\pm$ 4 & 11.38 $\pm$ 0.04 & 29.5 $\pm$ 1.0\\
2006 July 17 &     & Yes & \Kp  & 06:45:29 & 09:54:03 & 71 &  28 & 0.28 $\pm$ 0.05 & 65 $\pm$ 6 & 15.73 $\pm$ 0.09 &  6.8 $\pm$ 0.5\\
             &     &     & \Lp  & 06:46:41 & 09:52:50 & 69 &  30 & 0.65 $\pm$ 0.04 & 82 $\pm$ 2 & 12.85 $\pm$ 0.07 &  7.6 $\pm$ 0.5\\
2005 July 16 & NGS & No  & \Lp  & 07:25:31 & 09:44:02 & 26 &  60 & 0.50 $\pm$ 0.05 & 85 $\pm$ 2 & 13.00 $\pm$ 0.10 &  6.6 $\pm$ 0.6\\
             &     &     & \Ms  & 07:17:21 & 09:39:49 & 39 & 120 & 0.53 $\pm$ 0.11 & 99 $\pm$ 1 & 12.32 $\pm$ 0.25 &  7.3 $\pm$ 1.7\\
\enddata

\tablenotetext{a}{Uncertainties listed are the relative
  uncertainties obtained from the RMS flux density variations
  of non-variable stars of comparable magnitude.}

\tablenotetext{b}{Dereddened flux densities were obtained from
  the apparent magnitudes by assuming an extinction law of
  $A_{H}/A_V$=0.1771, $A_{K^\prime}/A_V$=0.1108,
  $A_{L^\prime}/A_V$=0.0540, $A_{M_{s}}/A_V$=0.0504
  \citep{moneti01}, a visual extinction of 29 $\pm$ 1
  (S.~A. Wright et al. 2007, in preparation), and flux zero
  points of 1050, 686, 249, 163 Jy for $H$, $K^\prime$,
  $L^\prime$, $M_{s}$, respectively \citep{tokunaga05}.}

\label{nirobstable}
\end{deluxetable}

%% file: tab2.tex
\begin{deluxetable*}{cccc}
\tabletypesize{\scriptsize}
\tablecaption{Summary of Spectral Indices for Sgr A*}
\tablewidth{0pt}
\tablehead{
\colhead{Date} &
\colhead{$\alpha_{H-K'}$} &
\colhead{$\alpha_{K'-L'}$} &
\colhead{$\alpha_{L'-M_s}$}
}
\startdata
2004 July 26                  &                                       & -0.68$\pm$(0.36$\pm$0.07)$\pm$[0.14] &                                      \\
2005 July 16                  &                                       &                                      & -0.87$\pm$(0.70$\pm$0.58)$\pm$[0.57] \\
2005 July 31 & -0.98$\pm$(0.15$\pm$0.14)$\pm$[0.27]\tablenotemark{a}  & -0.62$\pm$(0.02$\pm$0.21)$\pm$[0.14] &                                      \\
2006 May 02  &  0.00$\pm$(0.46$\pm$0.39)$\pm$[0.27]\tablenotemark{a}  & -0.42$\pm$(0.03$\pm$0.38)$\pm$[0.14] &                                      \\
2006 July 16                  &                                       & -0.51$\pm$(0.03$\pm$0.14)$\pm$[0.14] &                                      \\
\cline{1-4}
Weighted Average\tablenotemark{b}& -0.88$\pm$(0.19)$\pm$[0.27]        & -0.55$\pm$(0.11)$\pm$[0.14]          &  -0.87$\pm$(0.91)$\pm$[0.57]  \\
\enddata

\tablecomments{The three sources of uncertainty for each value
  are the uncertainty in the weighted mean of the individual
  measurements, the uncertainty due to the color excess
  correction technique, and the systematic uncertainty from
  the uncertainties in the assumed spectral indices for the
  calibration stars, respectively. We leave them separate here
  to emphasize the stability of the spectral index within each
  observation epoch (as evidenced by the first
  uncertainty). The third uncertainty (presented in square
  brackets) is a systematic effect that is the same for each
  measurement and, thus, does not go down with measurements
  made over multiple years. (See Appendix for details.)}

\tablenotetext{a}{After correction for background
  contamination.}

\tablenotetext{b}{The first two uncertainties for each
  date have been added in quadrature before taking the
  weighted average of each color pair and the first
  uncertainty listed here is the resulting uncertainty of the
  weighted mean. }
\label{indices}

\end{deluxetable*}

%% file: tab3.tex
\begin{deluxetable}{cccccccccc}
\tabletypesize{\scriptsize}
\tablecaption{Nearby Stars Used to Determine Color Excesses}
\tablewidth{0pt}
\tablehead{
\colhead{} &
\colhead{$K^\prime$} &
\colhead{Spectral} &
\colhead{2004 July} &
\multicolumn{3}{c}{2005 July} &
\multicolumn{2}{c}{2006 May} &
\colhead{2006 July} \\
\cline{5-7} \cline{8-9}
\colhead{Name} &
\colhead{(mag)}&
\colhead{Type\tablenotemark{a}} &
\colhead{E($K^\prime-L^\prime$)\tablenotemark{b}} &
\colhead{E($H-K^\prime$)} &
\colhead{E($K^\prime-L^\prime$)} &
\colhead{E($L^\prime-M_s$)} &
\colhead{E($H-K^\prime$)} &
\colhead{E($K^\prime-L^\prime$)} &
\colhead{E($K^\prime-L^\prime$)} 
}
\startdata
S0-1  & 14.52$\pm$0.02 & B0-2 &               & 2.10$\pm$0.02 & 1.60$\pm$0.02 &-0.02$\pm$0.11 & 2.05$\pm$0.04 & 1.60$\pm$0.13 & 1.62$\pm$0.07\\
S0-2  & 13.98$\pm$0.02 & B0   & 1.55$\pm$0.04 & 2.10$\pm$0.05 & 1.46$\pm$0.02 &-0.23$\pm$0.13 &               &               &              \\
S0-3  & 14.30$\pm$0.02 & B0-2 &               & 2.15$\pm$0.07 & 1.46$\pm$0.02 &               & 2.14$\pm$0.20 & 1.55$\pm$0.10 & 1.41$\pm$0.01\\
S0-4  & 14.20$\pm$0.02 & B0-2 &               & 2.14$\pm$0.02 & 1.55$\pm$0.03 & 0.06$\pm$0.08 & 2.07$\pm$0.06 & 1.69$\pm$0.10 & 1.55$\pm$0.04\\
%S0-5 & 14.84$\pm$0.03 & B0-2 &               & 2.23$\pm$0.07 & 1.61$\pm$0.03 &               &               &               & 1.24$\pm$0.18\\
S0-19 & 15.24$\pm$0.01 & B4-9 & 1.54$\pm$0.09 & 2.10$\pm$0.04 & 1.50$\pm$0.04 &               & 1.98$\pm$0.14 & 1.32$\pm$0.15 & 1.44$\pm$0.03\\
S0-20 & 15.64$\pm$0.03 & B4-9 & 1.62$\pm$0.21 & 2.04$\pm$0.04 & 1.23$\pm$0.10 &               &               &               &              \\ 
S0-26 & 15.12$\pm$0.02 & B4-9 &               & 2.11$\pm$0.05 & 1.63$\pm$0.04 &               & 1.86$\pm$0.22 & 1.11$\pm$0.41 & 1.56$\pm$0.02\\
\cline{1-10}
\multicolumn{3}{c}{Average Color Excess} & 1.55$\pm$0.04 &
2.11$\pm$0.04&1.53$\pm$0.13&-0.02$\pm$0.15& 2.05$\pm$0.11 & 1.56$\pm$0.24 &1.45$\pm$0.09\\
\multicolumn{3}{c}{Spectral Index Uncertainty\tablenotemark{c}} & 0.07 &
0.14&0.21&0.58& 0.39& 0.38 &0.14\\
\enddata

\tablenotetext{a}{From \citet{ghez03,eisen05}.}
\tablenotetext{b}{From reported values in \citet{ghez05lgs}.}
\tablenotetext{c}{Propagated from average color excess uncertainty.}
\label{compstars}
\end{deluxetable}